\documentclass[twocolumn,floats,floatfix, aps, prl,superscriptaddress,english]{revtex4-1}
\usepackage{amssymb}
\usepackage{graphicx}
\usepackage{amsmath}
\usepackage[latin9]{inputenc}
\usepackage{array}
\usepackage{multirow}
\usepackage{color}
\usepackage{esint}
\usepackage{bm}
\usepackage{color}
\usepackage{bbm}
\usepackage{hyperref}
\usepackage{babel}
\usepackage{titlesec}
\usepackage{graphicx,amssymb,color}
\usepackage{epsfig}
\usepackage{epstopdf}
\usepackage{mathtools}
\usepackage{soul}
\usepackage[dvipsnames]{xcolor}
\usepackage{tikz}

\hypersetup{colorlinks=true, linkcolor=black, citecolor=black, urlcolor=black}

\begin{document}

\setcounter{equation}{0} \setcounter{figure}{0}
\setcounter{table}{0} \setcounter{page}{1} \makeatletter

\title{Effective statistical fringe removal algorithm \\
	for high-sensitivity imaging of ultracold atoms}

\author{Bo Song}
\email{bsong@connect.ust.hk}
\affiliation{Department of Physics, The Hong Kong University of Science and Technology,\\ Clear Water Bay, Kowloon, Hong Kong, China}
\author{Chengdong He}
\affiliation{Department of Physics, The Hong Kong University of Science and Technology,\\ Clear Water Bay, Kowloon, Hong Kong, China}
\author{Zejian Ren}
\affiliation{Department of Physics, The Hong Kong University of Science and Technology,\\ Clear Water Bay, Kowloon, Hong Kong, China}
\author{Entong Zhao}
\affiliation{Department of Physics, The Hong Kong University of Science and Technology,\\ Clear Water Bay, Kowloon, Hong Kong, China}
\author{Jeongwon Lee}
\affiliation{Department of Physics, The Hong Kong University of Science and Technology,\\ Clear Water Bay, Kowloon, Hong Kong, China}
\affiliation{Institute for Advanced Study, The Hong Kong University of Science and Technology,\\ Clear Water Bay, Kowloon, Hong Kong, China}
\author{Gyu-Boong Jo}
\email{ gbjo@ust.hk}
\affiliation{Department of Physics, The Hong Kong University of Science and Technology,\\ Clear Water Bay, Kowloon, Hong Kong, China}

\date{\today}

\begin{abstract}
High-sensitivity imaging of ultracold atoms is often challenging when interference patterns are imprinted on the imaging light. Such image noises result in low signal-to-noise ratio and limit the capability to extract subtle physical quantities. Here we demonstrate an advanced fringe removal algorithm for absorption imaging of ultracold atoms, which efficiently suppresses unwanted fringe patterns using a small number of sample images without taking additional reference images. The protocol is based on an image decomposition and projection method with an extended image basis. We apply this scheme to raw absorption images of degenerate Fermi gases for the measurement of atomic density fluctuations and temperatures. The quantitative analysis shows that image noises can be efficiently removed with only tens of reference images, which manifests the efficiency of our protocol. Our algorithm would be of particular interest for the quantum emulation experiments in which several physical parameters need to be scanned within a limited time duration.\end{abstract}

\maketitle
\newpage

\section{Introduction}
Ultracold atomic systems have emerged as tunable experimental platforms ranging from an atomic clock and interferometer~\cite{Ludlow:2015ks} to a quantum emulator~\cite{Bloch:2012jy}. These advancements are enabled by the exceptional precision in the control of experimental conditions but the accurate detection of atoms is a prerequisite for achieving such high-precision measurement and control. To suppress the systematic error in the detection of atoms, one often takes and averages a sufficient number $N$ of images, which results the noise being scaled as $1/\sqrt{N}$. Such statistical averaging, however, requires a high repetition rate of data acquisition. Here we present an efficient protocol for image processing, which requires only a small number of images to statistically suppress the fringe patterns and enhance the signal-to-noise ratio (SNR). This method is of particular interest for the quantum gas experiments in which several physical parameters need to be scanned within a limited time duration. The protocol described here has been already employed to perform high-sensitivity measurements with ultracold atoms~\cite{contact,he2020}.



Various statistical techniques have been proposed and demonstrated to suppress noise patterns and to improve the SNR in imaging of atoms. Fringe removal algorithms~\cite{li2007reduction,ockeloen2010detection,niu2018optimized} have been widely used to improve the detection of small atom numbers~\cite{ockeloen2010detection}. Moreover, statistical analysis such as principle component analysis (PCA) and  independent component analysis (ICA)~\cite{segal2010revealing,dubessy2014imaging,maciej2016fringe,shioya2017fringe}, and advanced nonlinear machine learning algorithms ~\cite{rem2018identifying,cao2019extraction} have been implemented to extract spatial or temporal information for a given data set. For example, the quantum phase transition~\cite{rem2018identifying} or the collective excitations~\cite{dubessy2014imaging} were precisely investigated with those methods. 



In this work, we propose and demonstrate an efficient fringe removal protocol for high-sensitivity absorption imaging of ultracold atoms. Our protocol is based on statistical image decomposition and projection methods using the data images as a basis set and compensating for unwanted fringes. Different from the previous works~\cite{ockeloen2010detection} and the original ideas~\cite{erhard2004experimente,kronjager2007coherent}, we extend the number of the basis set based on the systematic defect of the imaging system, being confirmed by PCA, and show that a sufficiently high SNR can be achieved with a small number of images. We quantitatively demonstrate the enhancement of image quality by investigating atomic density fluctuations, power spectrum of spatial Fourier transform and thermometry with degenerate fermions. This method not only reduces the experimental duration of taking images, but also provides a sufficient image quality for further image data analysis. In our recent experiment~\cite{contact}, for example, the fringe removal protocol allowed us to examine the extremely low optical density (OD) regime which cannot be accurately analyzed without fringe removal~\cite{contact}.



\section{Method}
\subsection{Protocol for removing fringes}
When imaging atoms, fringes on the imaging light beam can be induced by various sources such as, the diffraction from optical elements, or the interference between adjacent optical surfaces. In typical absorption imaging of cold atoms, those fringe patterns emerge in images taken by the CCD camera. These images consist of an absorption image with atoms $I^{w}$, a reference image without atoms $I^{wo}$ and a dark image $I^{dark}$, which results in an optical density (OD) image  $I^{OD}=\ln(\frac{I^{wo}-I^{dark}}{I^{w}-I^{dark}})$. Note that this is valid for a small saturation parameter $I_0/I_{sat}\ll 1$, where $I_0$ and $I_{sat}$ are the imaging light intensity and the saturation intensity, respectively. For high intensity imaging at $I_0/I_{sat}\gtrsim 1$, $I^{OD}$ should be corrected by a linear term $I^{wo}-I^{w}$ \cite{Reinaudi07,hueck2017calibrating}. Ideally, absorption imaging technique is immune to any fringe if the first two images $I^{w}$ and $I^{wo}$ contain the same pattern of fringes appearing at the same position and the reference image normalizes intensity variation of the probe beam. Nevertheless, the experimental imperfections including the vibration of light beam lead fringe patterns in $I^{w}$ and $I^{wo}$ relatively displaced, yielding unwanted patterns shown on the final OD image. Therefore the key requirement for fringe removal processing is to find the matched pair of absorption ($I^{w}$) and reference ($I^{wo}$) images, where all background patterns are identical except the region containing the atomic signal. Typically, a set of multiple reference images under identical conditions are taken in order to form a reference basis ($I^{wo}$). Then an image with atom $I^{w}$ is projected on this basis with a set of coefficients, and a corrected reference image is composed of the ${I^{wo}}$ weighted by the coefficients~\cite{ockeloen2010detection}.

\begin{figure}
	\centering
	\includegraphics[width=1\linewidth]{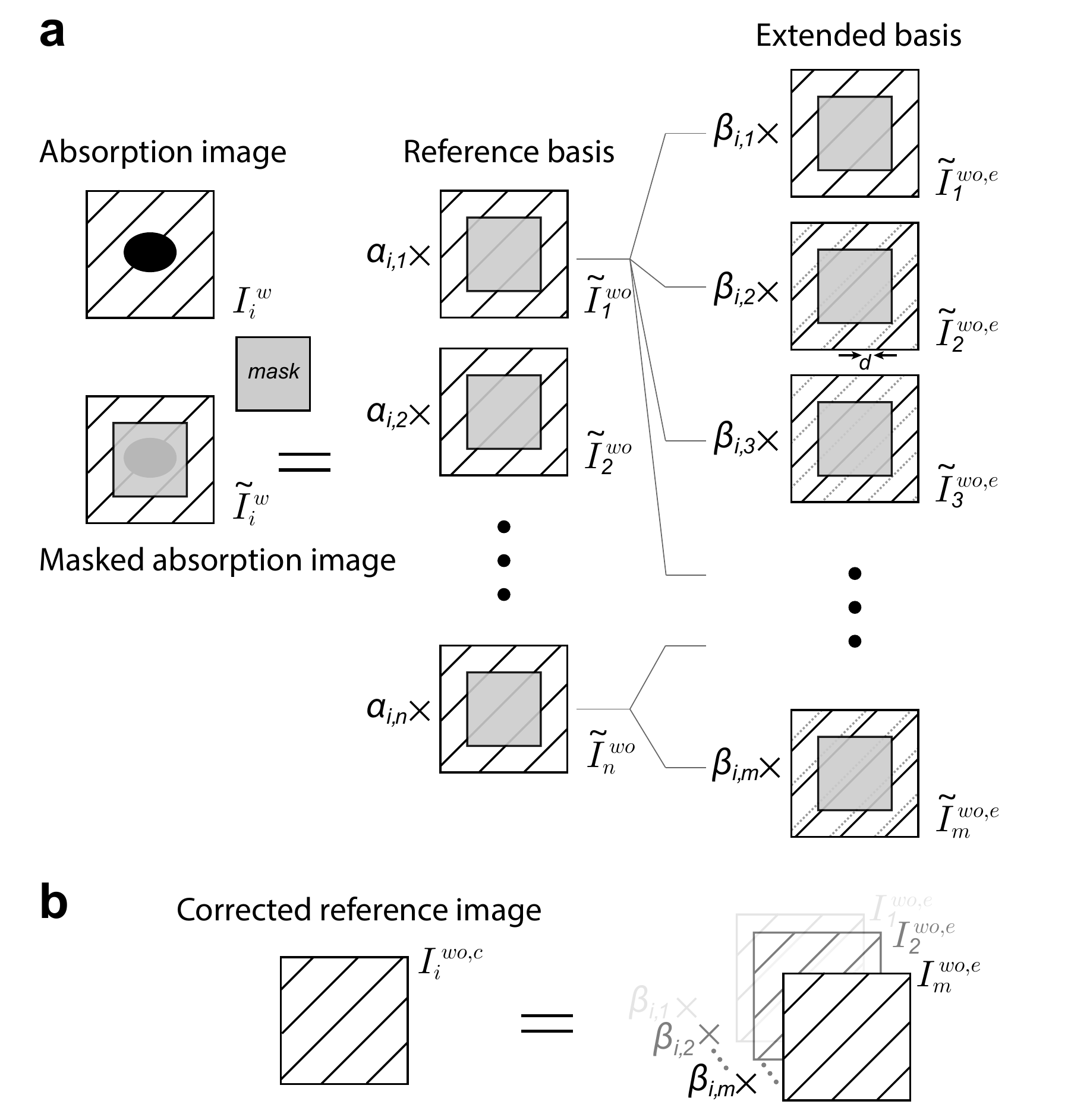}
	\caption{\textbf{Protocol of fringe removal algorithm} {The key of our fringe removal protocol is to reconstruct the corrected reference image $I^{wo,c}_i$, on which fringes would match their counterparts on the absorption image $I^{w}_i$. The procedure consists of {\bf{a}} decomposition and {\bf{b}} composition. We begin to exclude the atomic signal by masking both absorption $I^{w}_i$ and reference $I^{wo}_i$ images, resulting in $\tilde{I}^{w}_i$ and $\tilde{I}^{wo}_i$ respectively. Without extending a basis set, the masked image $\tilde{I}^{w}_i$ is decomposed into the basis set $\{\tilde{I}^{wo}_i \}$ as $\tilde{I}^{w}_i = \sum_j \alpha_{i,j} \tilde{I}^{wo}_i $ with corresponding coefficients $\alpha_{i,j}$. Then, the corrected reference image for $I^{w}_i$ is reconstructed as $I^{wo,c}_i=\sum_j \alpha_{i,j} I^{wo}_{j}$. The key improvement in this work is that a new extended basis set is obtained by shifting an original set $\{ \tilde{I}^{wo}\}$ by $-d, -(d-1), ..., d$ pixels and used for image decomposition. Then the masked image $\tilde{I}^{w}_i$ is decomposed on the extended set as $\tilde{I}^{w}_i = \sum_j \beta_{i,j} \tilde{I}^{wo,e}_{j}$. Finally, the corrected reference image is reconstructed as $I^{wo,c}_i=\sum_j \beta_{i,j} I^{wo,e}_{j}$.}}
	\label{Fig1_Schematic}
\end{figure}

Our fringe removal protocol consists two steps, the image decomposition and composition with an extended basis set, as depicted in Fig.~\ref{Fig1_Schematic}. The performance of the fringe removal highly relies on the completeness of the basis set formed from reference images ($I^{wo}$), which is directly associated with the compensation to fringe patterns in $I^{w}$ by a corrected reference image. Ideally, the first two images should contain the same pattern except the region containing atoms. To avoid the effect from atoms, the algorithm starts with masking the region containing atoms, and has the certain masked absorption images $\tilde{I}^{w}_{i}$ ($i=1...n$), and a set of masked reference images ${\tilde{I}^{wo}_{j}} (j=1...n)$. In the previous method~\cite{ockeloen2010detection},  each image $\tilde{I}^{w}_{i}$ is decomposed into the masked basis $\{\tilde{I}^{wo}_{j}\}$ and the coefficients are ${\alpha_{i,j}}$. Then the corrected reference image can be reconstructed as the sum of ${I}^{w}_{j}$ weighted by ${\alpha_{i,j}}$, which minimizes the least square difference between the absorption and the corrected reference image. However, if the pattern is relatively shifted during the imaging process due to experimental imperfections, the fringe pattern sometimes cannot be reproduced by the basis set $\{I^{wo}\}$. Consequently, the corrected image cannot match with the certain absorption image, $I^{w}_{i}$. In principle, we can increase the number of basis by acquiring sufficient sample images, which would allow us to compensate for such shifted fringes. However, this would inevitably take more time for data acquisition, which is not ideal for quantum emulation experiments required to be performed within a limited time duration. Compared with the previous method without an extended basis, our improved protocol not only eliminates the background fringe with a small number of images, but also delivers superior performance.

Here, the improved method is to spatially shift images in the original set few pixels both horizontally and vertically, resulting in an extended set $\{\tilde{I}^{wo,e}_{j}\}$. The coefficient $\beta_{i,j}$ is determined by the decomposition of $I^w_i$ into this new basis set. The autocorrelation of the basis $C$ is calculated as,
\begin{equation}
	C_{j,k} = \sum_{x,y} \tilde{I}^{wo,e}_{j}(x,y) \cdot \tilde{I}^{wo,e}_{k}(x,y)
\end{equation}
where $(x,y)$ denotes the x-y position of the image. The projection of masked absorption image $\tilde{I}^{w}_{i}$ on the basis, $P$ is as follows,
\begin{equation}
	P_{i, k} = \sum_{x,y} \tilde{I}^{w}_{i}(x,y) \cdot \tilde{I}^{wo,e}_{j}(x,y)
\end{equation}
The coefficient $\beta$ is therefore extracted by solving the matrix equation,
\begin{equation}
	P \cdot \beta = C
\end{equation}
The final corrected reference image for certain $i$ is composed as,
\begin{equation}
	I^{wo,c}_{i} = \sum_{j}\beta_{i,j} I^{wo,e}_{i}
\end{equation}
where $I^{wo,e}_{i}$ is the unmasked extended basis for $i=1,...,m=(2d+1)^2$, which finally minimizes the least square difference $\sum_{x,y} (\tilde{I}^{w}_{i}$-$\tilde{I}^{wo,c}_{i})^2$. To be noted, this algorithm corrects not only the fringe patterns but also the intensity difference of the imaging light between reference and absorption images. In addition, this algorithm recovers the corrected reference image $I_i^{wo,c}$ based on a masked absorption image set $\{\tilde{I}_i^{w}\}$ instead of $\{I_i^{w}\}$. This allows to apply the fringe removal protocol to a series of data images taken in time, for example, in the measurement of collective modes in quantum gases~\cite{he2020}.

\subsection{Construction of extended basis}
\begin{figure}
	\centering
	\includegraphics[width=0.9\linewidth]{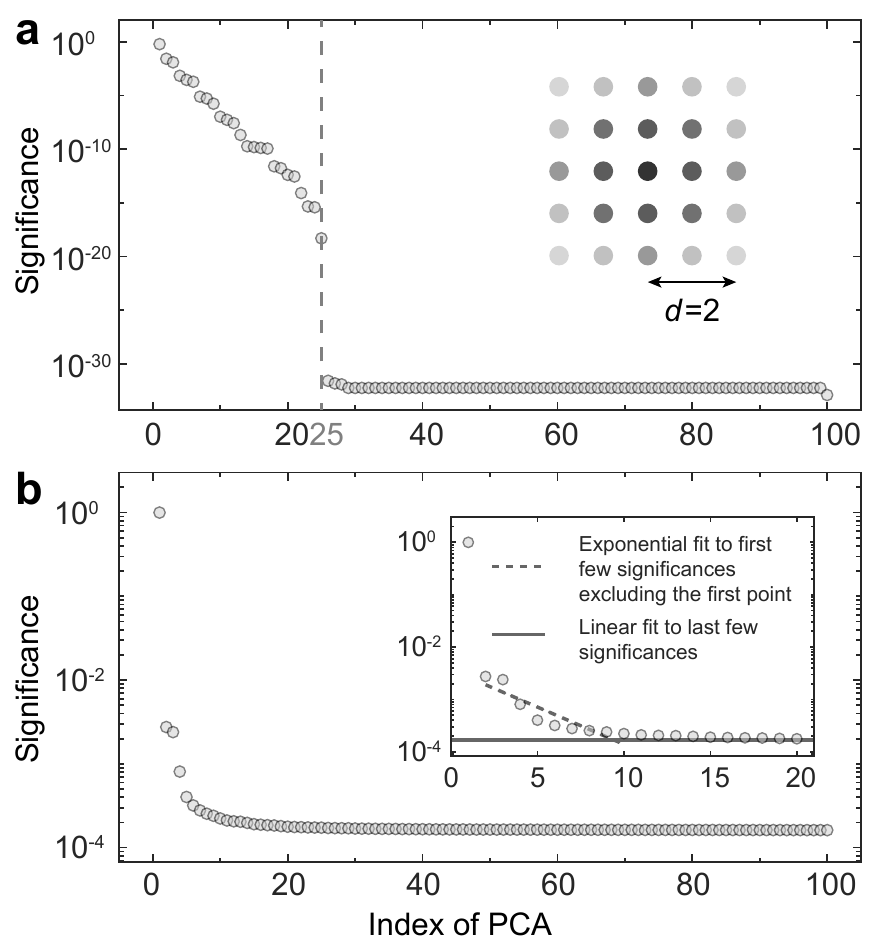}
	\caption{\textbf{Rank of simulated and measured set}. \textbf{a}, the rank of a simulated set. The simulated data set consists of 100 images generated by randomly shifting image fringes both horizontally and vertically within $d=2$ pixels. The significance, the eigen value from the result of PCA, reflecting the importance of each principal image, is plotted as a function of ranked principal images of the simulated data set. The result shows that only the first 25 images are significant, which is consistent with the rank of the data set, $(2d+1)^2=25$. \textbf{b}, the rank of the experimental set. The first few images are significant, and the shift d=1 is sufficient for the extended basis in the fringe removal algorithm. Inset shows the result of the exponential fit of the first 10 principal images in dashed line and the final saturated value in solid line by a linear fit to the last 10 principal images.}
	\label{Fig2_SetShift}
\end{figure}

To run the fringe removal protocol in a cost effective way,  it is important to construct the extended basis set of reference images with the minimal shift $d$ in units of pixels. We run PCA for the set of images, and determine $d$ from the rank of the set. We test this protocol by analysing a set of 100 images numerically generated with two types of fringes (ring and linear fringes). In this set, we assume that all fringe patterns move together and randomly shift within $d=2$ pixels in both horizontal and vertical directions (see supplemental material). By applying PCA, we show the significance, which is the eigenvalue of each principal image (i.e. eigenvector), as a function of the index of the principal image in Fig.\ref{Fig2_SetShift}~(a). The significance {physically} stands for the dominance of principal images within the data set. The result clearly shows that only the first 25 principal images are significant, which is consistent with the rank of this data set $(2d +1)^2=25$.

In a real imaging system, fringes originate from different sources, which increases the degrees of freedom of patterns. For example, the total rank could be as large as $25^2=625$ if both patterns shift up to $d=2$ independently.  Note that it is important to physically suppress fringes (e.g. cleaning optics and avoiding light interference), which will minimize the computational complexity. We apply PCA to the reference images $\{ I^{wo} \}$ in our imaging system, and estimate the rank of the set. As shown in Fig.~\ref{Fig2_SetShift}(b), we find that the fringes shift within 1 pixel in our imaging system. Therefore, the extended basis with $d=1$ should be sufficient for the fringe removal algorithm.

\begin{figure}
	\centering
	\includegraphics[width=0.85\linewidth]{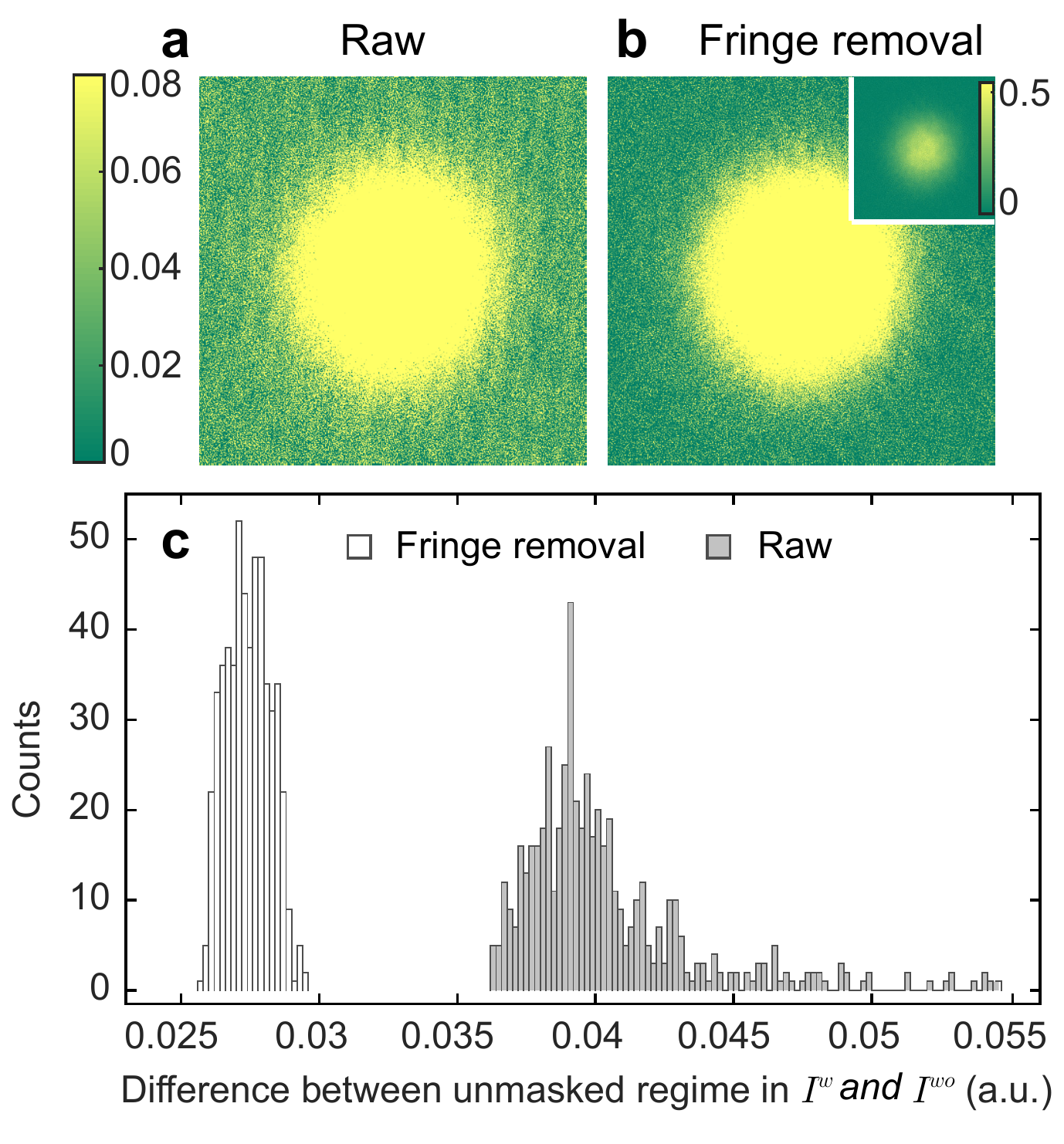}
	\caption{\textbf{Comparison between absorption imaging with and without fringe removal.} \textbf{a}, a typical OD image shows a background fringe. \textbf{b}, the same OD image with fringe removal in the same colour scale as \textbf{a}. Inset is the OD image with fringe removal in a full colour scale. \textbf{c}, difference between absorption and reference images in the absorption imaging. Ideally, these two images should be identical except the area containing atoms. The difference between these two masked images of absorption images reflects cleanness of the background in the final OD image. We repeat the same absorption imaging $\sim$500 times, and statistically count and compare the difference between the case with (open) and without (filled) fringe removal. The difference is substantially reduced by fringe removal.}
	\label{Fig3_ImgResult}
\end{figure}

\section{Result}
In Fig.~\ref{Fig3_ImgResult}, we show a typical OD image ($I^{OD}$) of atoms with and without applying the fringe removal protocol. Here, a thermal Fermi gas of $1 \times 10^5$ $^{173}$Yb atoms at 4$\mu$K is ballistically expanded for 4~ms, followed by absorption imaging with a resonant atomic transition light. To remove fringes being present in the current system as shown in Fig.~\ref{Fig3_ImgResult}(a), we apply our statistical fringe removal protocol with the basis set $\{I^{wo}\}$ of 30 reference images. We extend the number of basis by a factor of $(2d+1)^2=9$ shifting the basis image in the horizontal and vertical directions within $d=1$. Fig.~\ref{Fig3_ImgResult}(b) is a single-shot result of absorption imaging with fringes suppressed by the proposed protocol. To characterize the image quality quantitatively, we monitor the difference between $I^{w}$ and $I^{wo}$ in the background region, $\sum_{x,y} m_{x,y}\lvert I^{w}-I^{wo} \rvert$ where $m_{x,y}$ is a mask function excluding the atomic signal ($m_{x,y}=0$). Fig.~\ref{Fig3_ImgResult}(c) shows the histogram of difference from 500 images of atoms. We find that the fringe removal process not only reduces the difference between $I^{w}$ and $I^{wo}$ but also minimizes the systematic fluctuation in OD (i.e. the width of the distribution).


\begin{figure}[t]
	\centering
	\includegraphics[width=1\linewidth]{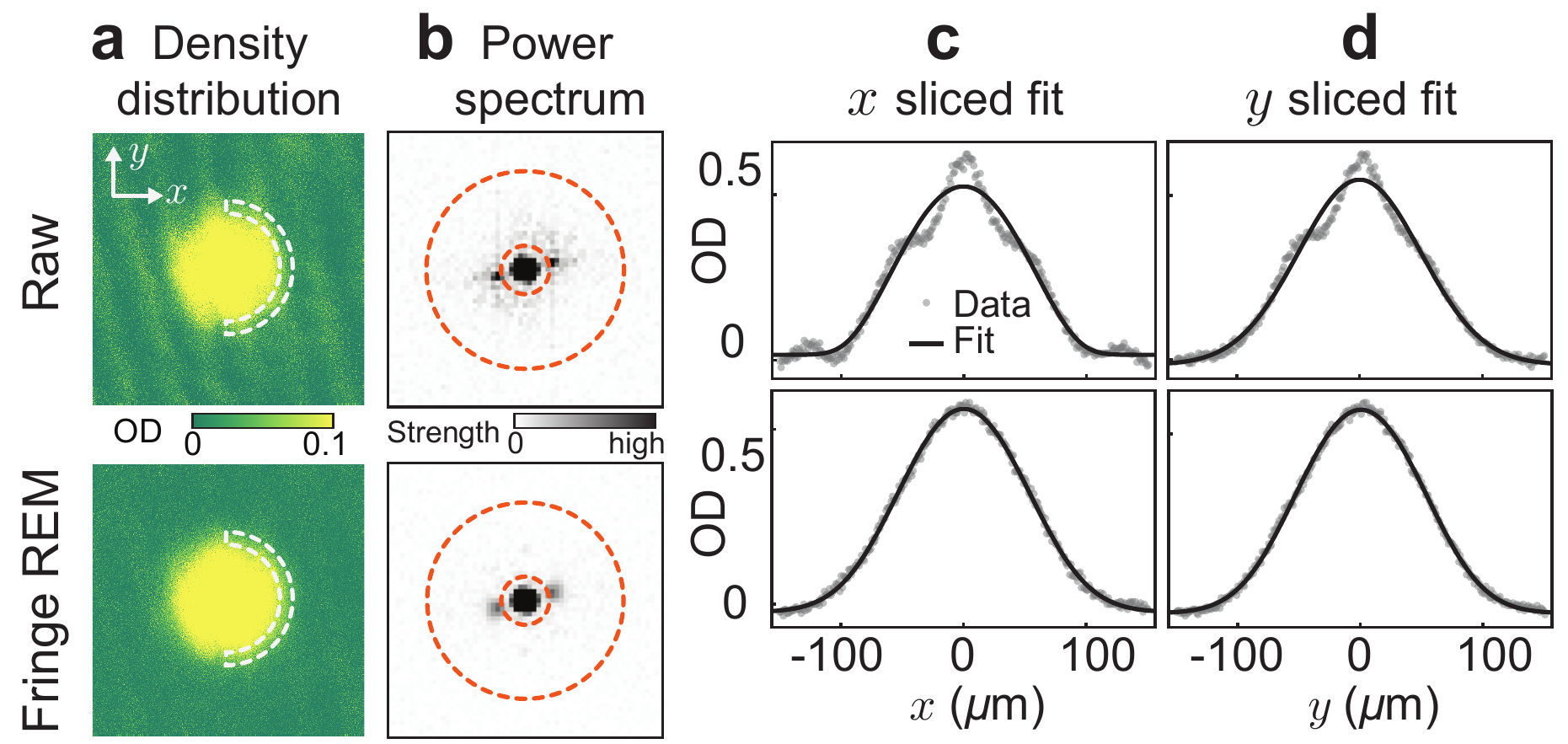}
	\caption{\textbf{Quantitative analysis of absorption imaging}. To quantitatively characterize the imaging, we engineer three features from the images of a ballistically expanded Fermi gas. \textbf{a}, the atom density variance enlarged by the fringes is extracted from the standard derivation of the OD in the white half-ring-circle dashed region. \textbf{b}, based on the power spectrum of the spatial Fourier transform of the image, the fringe strength is defined as the integrated spectral power in the annular orange dashed region. \textbf{c} and \textbf{d} are the slice profile along the $x$ and $y$ direction respectively. The variance of the temperature difference between these two directions is intensified by and thus reflects the fringes.}
	\label{Fig4_Features}
\end{figure}

{\bf Quantitative analysis of absorption imaging} To further illustrate the performance of the proposed scheme, we apply the protocol to raw absorption images of ultracold atoms taken in experiments~\cite{Song:2016ep}. We quantitatively (1) extract the fluctuation of OD (i.e. atomic density $n(k)$) at the certain momentum $k$, (2) measure the power spectrum of the spatial Fourier transform and (3) perform the temperature measurement as discussed later, which all indicate that the fringe removal protocol allows us to precisely extract subtle physical quantities.

\begin{figure}
	\centering
	\includegraphics[width=1.0\linewidth]{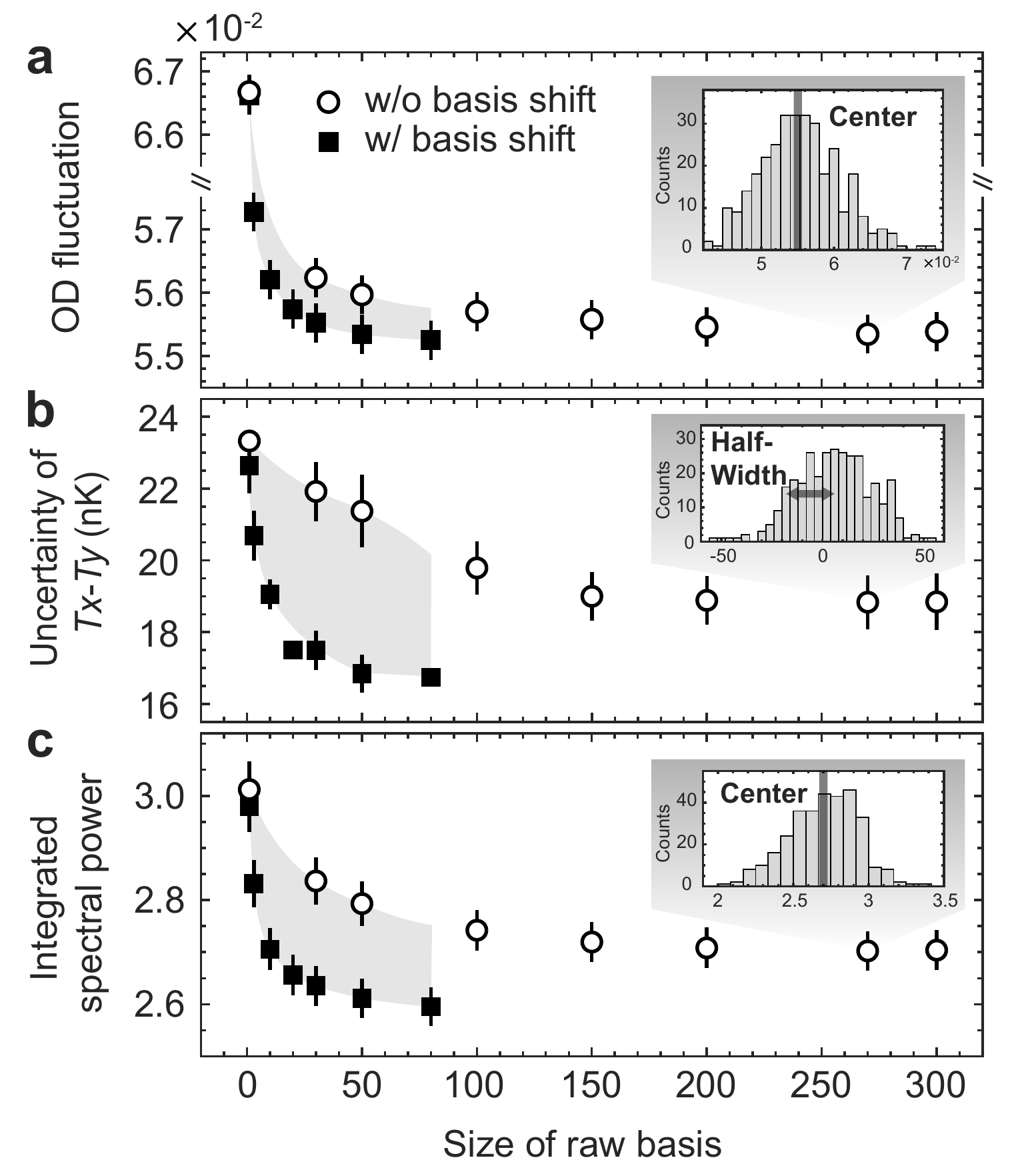}
	\caption{\textbf{Fringe removal performance influenced by the basis size}. Three features reflecting the strength of fringes are plotted as a function of the number of images used as the basis in the fringe removal algorithm. {\bf a}, OD fluctuation is determined by the central value of the distribution of OD standard deviation. {\bf b}, temperature difference uncertainty from the width of the distribution of the temperature difference $T_x - T_y$. {\bf c}, the central value of the distribution of fringe strength obtained from image Fourier transform. Without additional basis extension, the performance becomes saturated at 150 images, whereas the performance with $d$=1 shift starts to become saturated at 30 images. The performance of 30 images basis with shift is comparable to, even better than the one of 150 image basis without basis extension.}
	\label{FigX_Features}
\end{figure}

In Fig.~\ref{Fig4_Features}, we monitor a spin-polarized degenerate Fermi gas of $1 \times 10^5$ $^{173}$Yb atoms at a temperature $T=70$~nK prepared in an optical dipole trap where $T_F=$~200~nK is the Fermi temperature~\cite{Song:2016ep}. The sample is ballistically expanded for 20~ms before the absorption imaging, which minimizes the trap effect and ensures the isotropic momentum distribution after an expansion. We first examine the atomic density $n(k)$ at the constant momentum $k$ by measuring the standard deviation of optical density within the half-annular region. The variance of atomic density within the region reflects the strength of fringe patterns. Here, photon shot noise effectively contributes a variance around 0.03 in the OD fluctuation in Fig.~\ref{FigX_Features}(a). Secondly, we obtain the power spectrum of the spatial Fourier transform of the OD image. We characterize the strength of the background fringe pattern by integrating the spectral power in the annular region indicated by the dashed lines. Finally, we test the thermometry based on the atomic distribution with and without the presence of fringe patterns. We examine atomic profiles sliced along the $x$ and $y$ directions and extract temperatures $T_x$ and $T_y$, respectively, by Thomas-Fermi fits (see Fig.~\ref{Fig4_Features}). 


We now quantitatively examine the performance of fringe removal protocol with respect to the number of reference images. For the purpose, we repeat the same measurement $\sim$300 times in total and each OD image is processed by the fringe removal with the varying  size of the raw basis. The raw basis is chosen as a set of images taken consecutively near the analysed image. The standard deviation of atomic densities within the half annular region is calculated for each corrected OD image (see Fig.~\ref{Fig4_Features}(a)). For a given number of the raw basis, these standard deviations are statistically represented by a distribution of standard deviations like the inset of Fig.~\ref{FigX_Features}(a). Here, the center or the mean of the distribution reflects the background fringes. Without applying the fringe removal protocol, the fluctuation of the density $n(k)$ is relatively large as indicated by the grey area in Fig.~\ref{FigX_Features}. Here, we choose the statistical mean of the distribution as the density fluctuation. Without extending the basis set, the density fluctuation (open circles) decreases and becomes saturated after the basis size is larger than $\sim$150 images. Notably, it becomes quickly saturated around 30 images (filled squares) with the extended basis. In other words, our fringe removal is strikingly effective and only requires 30 images as the basis in our system in contrast to the previous methods using a few hundred images~\cite{ockeloen2010detection,erhard2004experimente,kronjager2007coherent}.

The effectiveness of our fringe removal protocol with an extended basis is further confirmed in Fig.~\ref{FigX_Features}(b,c). We examine the distribution of ($T_x-T_y$) and measure the width of the distribution (see the inset of Fig.~\ref{FigX_Features}(b)). The fringe removal protocol with an extended basis set (i.e. shifted by $d$=1) allows for more accurate thermometry than the previous method with a large number of reference images. In addition, we directly quantify the strength of background fringes from the power spectrum of the spatial Fourier transform of the OD image. Fig.~\ref{FigX_Features}(c) shows the effect of number of reference basis images on integrated power spectrum of fringes. It is clearly demonstrated that our protocol results in a faster decrease of the integrated power spectrum with a lower saturated value, compared to the fringe removal method without the basis extension. In addition, the background fringe can be more effectively suppressed with the basis extension resulting in smaller saturated spectral power. This suggests that our protocol not only reduces the measurement times, but also substantially enhances the image quality.

\section{Conclusion} We have developed an optimized fringe removal algorithm for absorption imaging technique, which can also be implemented into other imaging techniques such as phase contrast imaging. This method extends an image basis by shifting the basis which takes into account the possible mismatch between the absorption and reference images, often caused by mechanical vibrations in the imaging system. The shift depends on the drift of imaging system and is empirically determined by the PCA method. We have shown that an image can be largely improved by fringe removal with a small number of images. The protocol presented here has been recently implemented in Ref.~\cite{contact} wherein we examine the high-momentum atomic density in the time-of-flight distribution of ultracold fermions. Although we demonstrate the effectiveness of the fringe removal with extended basis with ultracold atoms, the protocol is generally applicable to any set of images taken in different systems.

\vspace{5pt}
\paragraph*{\bf Acknowledgement} G.-B. J. acknowledges the generous support from the Hong Kong Research Grants Council and the Croucher Foundation through  16311516, and 16305317, 16304918, 16306119, C6005-17G, N-HKUST601/17 and the Croucher Innovation grants respectively.

\bibliography{FringeRemoval}

\begin{thebibliography}{18}%
\makeatletter
\providecommand \@ifxundefined [1]{%
 \@ifx{#1\undefined}
}%
\providecommand \@ifnum [1]{%
 \ifnum #1\expandafter \@firstoftwo
 \else \expandafter \@secondoftwo
 \fi
}%
\providecommand \@ifx [1]{%
 \ifx #1\expandafter \@firstoftwo
 \else \expandafter \@secondoftwo
 \fi
}%
\providecommand \natexlab [1]{#1}%
\providecommand \enquote  [1]{``#1''}%
\providecommand \bibnamefont  [1]{#1}%
\providecommand \bibfnamefont [1]{#1}%
\providecommand \citenamefont [1]{#1}%
\providecommand \href@noop [0]{\@secondoftwo}%
\providecommand \href [0]{\begingroup \@sanitize@url \@href}%
\providecommand \@href[1]{\@@startlink{#1}\@@href}%
\providecommand \@@href[1]{\endgroup#1\@@endlink}%
\providecommand \@sanitize@url [0]{\catcode `\\12\catcode `\$12\catcode
  `\&12\catcode `\#12\catcode `\^12\catcode `\_12\catcode `\%12\relax}%
\providecommand \@@startlink[1]{}%
\providecommand \@@endlink[0]{}%
\providecommand \url  [0]{\begingroup\@sanitize@url \@url }%
\providecommand \@url [1]{\endgroup\@href {#1}{\urlprefix }}%
\providecommand \urlprefix  [0]{URL }%
\providecommand \Eprint [0]{\href }%
\providecommand \doibase [0]{http://dx.doi.org/}%
\providecommand \selectlanguage [0]{\@gobble}%
\providecommand \bibinfo  [0]{\@secondoftwo}%
\providecommand \bibfield  [0]{\@secondoftwo}%
\providecommand \translation [1]{[#1]}%
\providecommand \BibitemOpen [0]{}%
\providecommand \bibitemStop [0]{}%
\providecommand \bibitemNoStop [0]{.\EOS\space}%
\providecommand \EOS [0]{\spacefactor3000\relax}%
\providecommand \BibitemShut  [1]{\csname bibitem#1\endcsname}%
\let\auto@bib@innerbib\@empty
\bibitem [{\citenamefont {Ludlow}\ \emph {et~al.}(2015)\citenamefont {Ludlow},
  \citenamefont {Boyd}, \citenamefont {Ye}, \citenamefont {Peik},\ and\
  \citenamefont {Schmidt}}]{Ludlow:2015ks}%
  \BibitemOpen
  \bibfield  {author} {\bibinfo {author} {\bibfnamefont {A.~D.}\ \bibnamefont
  {Ludlow}}, \bibinfo {author} {\bibfnamefont {M.~M.}\ \bibnamefont {Boyd}},
  \bibinfo {author} {\bibfnamefont {J.}~\bibnamefont {Ye}}, \bibinfo {author}
  {\bibfnamefont {E.}~\bibnamefont {Peik}}, \ and\ \bibinfo {author}
  {\bibfnamefont {P.~O.}\ \bibnamefont {Schmidt}},\ }\href@noop {} {\bibfield
  {journal} {\bibinfo  {journal} {Reviews of Modern Physics}\ }\textbf
  {\bibinfo {volume} {87}},\ \bibinfo {pages} {637} (\bibinfo {year}
  {2015})}\BibitemShut {NoStop}%
\bibitem [{\citenamefont {Bloch}\ \emph {et~al.}(2012)\citenamefont {Bloch},
  \citenamefont {Dalibard},\ and\ \citenamefont
  {Nascimb{\`e}ne}}]{Bloch:2012jy}%
  \BibitemOpen
  \bibfield  {author} {\bibinfo {author} {\bibfnamefont {I.}~\bibnamefont
  {Bloch}}, \bibinfo {author} {\bibfnamefont {J.}~\bibnamefont {Dalibard}}, \
  and\ \bibinfo {author} {\bibfnamefont {S.}~\bibnamefont {Nascimb{\`e}ne}},\
  }\href@noop {} {\bibfield  {journal} {\bibinfo  {journal} {Nature Physics}\
  }\textbf {\bibinfo {volume} {8}},\ \bibinfo {pages} {267} (\bibinfo {year}
  {2012})}\BibitemShut {NoStop}%
\bibitem [{\citenamefont {Song}\ \emph {et~al.}(2019)\citenamefont {Song},
  \citenamefont {Yan}, \citenamefont {He}, \citenamefont {Ren}, \citenamefont
  {Zhou},\ and\ \citenamefont {Jo}}]{contact}%
  \BibitemOpen
  \bibfield  {author} {\bibinfo {author} {\bibfnamefont {B.}~\bibnamefont
  {Song}}, \bibinfo {author} {\bibfnamefont {Y.}~\bibnamefont {Yan}}, \bibinfo
  {author} {\bibfnamefont {C.}~\bibnamefont {He}}, \bibinfo {author}
  {\bibfnamefont {Z.}~\bibnamefont {Ren}}, \bibinfo {author} {\bibfnamefont
  {Q.}~\bibnamefont {Zhou}}, \ and\ \bibinfo {author} {\bibfnamefont {G.-B.}\
  \bibnamefont {Jo}},\ }\href@noop {} {\bibfield  {journal} {\bibinfo
  {journal} {arXiv preprint arXiv:1912.12105}\ } (\bibinfo {year}
  {2019})}\BibitemShut {NoStop}%
\bibitem [{\citenamefont {He}\ \emph {et~al.}(2020)\citenamefont {He},
  \citenamefont {Ren}, \citenamefont {Song}, \citenamefont {Zhao},
  \citenamefont {Lee}, \citenamefont {Zhang}, \citenamefont {Zhang},\ and\
  \citenamefont {Jo}}]{he2020}%
  \BibitemOpen
  \bibfield  {author} {\bibinfo {author} {\bibfnamefont {C.}~\bibnamefont
  {He}}, \bibinfo {author} {\bibfnamefont {Z.}~\bibnamefont {Ren}}, \bibinfo
  {author} {\bibfnamefont {B.}~\bibnamefont {Song}}, \bibinfo {author}
  {\bibfnamefont {E.}~\bibnamefont {Zhao}}, \bibinfo {author} {\bibfnamefont
  {J.}~\bibnamefont {Lee}}, \bibinfo {author} {\bibfnamefont {Y.-C.}\
  \bibnamefont {Zhang}}, \bibinfo {author} {\bibfnamefont {S.}~\bibnamefont
  {Zhang}}, \ and\ \bibinfo {author} {\bibfnamefont {G.-B.}\ \bibnamefont
  {Jo}},\ }\href@noop {} {\bibfield  {journal} {\bibinfo  {journal} {Physical
  Review Research}\ }\textbf {\bibinfo {volume} {2}},\ \bibinfo {pages}
  {012028} (\bibinfo {year} {2020})}\BibitemShut {NoStop}%
\bibitem [{\citenamefont {Li}\ \emph {et~al.}(2007)\citenamefont {Li},
  \citenamefont {Ke}, \citenamefont {Yan},\ and\ \citenamefont
  {Wang}}]{li2007reduction}%
  \BibitemOpen
  \bibfield  {author} {\bibinfo {author} {\bibfnamefont {X.}~\bibnamefont
  {Li}}, \bibinfo {author} {\bibfnamefont {M.}~\bibnamefont {Ke}}, \bibinfo
  {author} {\bibfnamefont {B.}~\bibnamefont {Yan}}, \ and\ \bibinfo {author}
  {\bibfnamefont {Y.}~\bibnamefont {Wang}},\ }\href@noop {} {\bibfield
  {journal} {\bibinfo  {journal} {Chinese Optics Letters}\ }\textbf {\bibinfo
  {volume} {5}},\ \bibinfo {pages} {128} (\bibinfo {year} {2007})}\BibitemShut
  {NoStop}%
\bibitem [{\citenamefont {Ockeloen}\ \emph {et~al.}(2010)\citenamefont
  {Ockeloen}, \citenamefont {Tauschinsky}, \citenamefont {Spreeuw},\ and\
  \citenamefont {Whitlock}}]{ockeloen2010detection}%
  \BibitemOpen
  \bibfield  {author} {\bibinfo {author} {\bibfnamefont {C.}~\bibnamefont
  {Ockeloen}}, \bibinfo {author} {\bibfnamefont {A.}~\bibnamefont
  {Tauschinsky}}, \bibinfo {author} {\bibfnamefont {R.}~\bibnamefont
  {Spreeuw}}, \ and\ \bibinfo {author} {\bibfnamefont {S.}~\bibnamefont
  {Whitlock}},\ }\href@noop {} {\bibfield  {journal} {\bibinfo  {journal}
  {Physical Review A}\ }\textbf {\bibinfo {volume} {82}},\ \bibinfo {pages}
  {061606} (\bibinfo {year} {2010})}\BibitemShut {NoStop}%
\bibitem [{\citenamefont {Niu}\ \emph {et~al.}(2018)\citenamefont {Niu},
  \citenamefont {Guo}, \citenamefont {Zhan}, \citenamefont {Chen},
  \citenamefont {Liu},\ and\ \citenamefont {Zhou}}]{niu2018optimized}%
  \BibitemOpen
  \bibfield  {author} {\bibinfo {author} {\bibfnamefont {L.}~\bibnamefont
  {Niu}}, \bibinfo {author} {\bibfnamefont {X.}~\bibnamefont {Guo}}, \bibinfo
  {author} {\bibfnamefont {Y.}~\bibnamefont {Zhan}}, \bibinfo {author}
  {\bibfnamefont {X.}~\bibnamefont {Chen}}, \bibinfo {author} {\bibfnamefont
  {W.}~\bibnamefont {Liu}}, \ and\ \bibinfo {author} {\bibfnamefont
  {X.}~\bibnamefont {Zhou}},\ }\href@noop {} {\bibfield  {journal} {\bibinfo
  {journal} {Applied Physics Letters}\ }\textbf {\bibinfo {volume} {113}},\
  \bibinfo {pages} {144103} (\bibinfo {year} {2018})}\BibitemShut {NoStop}%
\bibitem [{\citenamefont {Segal}\ \emph {et~al.}(2010)\citenamefont {Segal},
  \citenamefont {Diot}, \citenamefont {Cornell}, \citenamefont {Zozulya},\ and\
  \citenamefont {Anderson}}]{segal2010revealing}%
  \BibitemOpen
  \bibfield  {author} {\bibinfo {author} {\bibfnamefont {S.~R.}\ \bibnamefont
  {Segal}}, \bibinfo {author} {\bibfnamefont {Q.}~\bibnamefont {Diot}},
  \bibinfo {author} {\bibfnamefont {E.~A.}\ \bibnamefont {Cornell}}, \bibinfo
  {author} {\bibfnamefont {A.~A.}\ \bibnamefont {Zozulya}}, \ and\ \bibinfo
  {author} {\bibfnamefont {D.~Z.}\ \bibnamefont {Anderson}},\ }\href@noop {}
  {\bibfield  {journal} {\bibinfo  {journal} {Physical Review A}\ }\textbf
  {\bibinfo {volume} {81}},\ \bibinfo {pages} {053601} (\bibinfo {year}
  {2010})}\BibitemShut {NoStop}%
\bibitem [{\citenamefont {Dubessy}\ \emph {et~al.}(2014)\citenamefont
  {Dubessy}, \citenamefont {De~Rossi}, \citenamefont {Badr}, \citenamefont
  {Longchambon},\ and\ \citenamefont {Perrin}}]{dubessy2014imaging}%
  \BibitemOpen
  \bibfield  {author} {\bibinfo {author} {\bibfnamefont {R.}~\bibnamefont
  {Dubessy}}, \bibinfo {author} {\bibfnamefont {C.}~\bibnamefont {De~Rossi}},
  \bibinfo {author} {\bibfnamefont {T.}~\bibnamefont {Badr}}, \bibinfo {author}
  {\bibfnamefont {L.}~\bibnamefont {Longchambon}}, \ and\ \bibinfo {author}
  {\bibfnamefont {H.}~\bibnamefont {Perrin}},\ }\href@noop {} {\bibfield
  {journal} {\bibinfo  {journal} {New Journal of Physics}\ }\textbf {\bibinfo
  {volume} {16}},\ \bibinfo {pages} {122001} (\bibinfo {year}
  {2014})}\BibitemShut {NoStop}%
\bibitem [{\citenamefont {Trusiak}\ \emph {et~al.}(2016)\citenamefont
  {Trusiak}, \citenamefont {S{\l}u\.{z}ewski},\ and\ \citenamefont
  {Patorski}}]{maciej2016fringe}%
  \BibitemOpen
  \bibfield  {author} {\bibinfo {author} {\bibfnamefont {M.}~\bibnamefont
  {Trusiak}}, \bibinfo {author} {\bibfnamefont {{\L}.}~\bibnamefont
  {S{\l}u\.{z}ewski}}, \ and\ \bibinfo {author} {\bibfnamefont
  {K.}~\bibnamefont {Patorski}},\ }\href {\doibase 10.1364/OE.24.004221}
  {\bibfield  {journal} {\bibinfo  {journal} {Opt. Express}\ }\textbf {\bibinfo
  {volume} {24}},\ \bibinfo {pages} {4221} (\bibinfo {year}
  {2016})}\BibitemShut {NoStop}%
\bibitem [{\citenamefont {Shioya}\ \emph {et~al.}(2017)\citenamefont {Shioya},
  \citenamefont {Shimoaka},\ and\ \citenamefont {Hasegawa}}]{shioya2017fringe}%
  \BibitemOpen
  \bibfield  {author} {\bibinfo {author} {\bibfnamefont {N.}~\bibnamefont
  {Shioya}}, \bibinfo {author} {\bibfnamefont {T.}~\bibnamefont {Shimoaka}}, \
  and\ \bibinfo {author} {\bibfnamefont {T.}~\bibnamefont {Hasegawa}},\
  }\href@noop {} {\bibfield  {journal} {\bibinfo  {journal} {Analytical
  Sciences}\ }\textbf {\bibinfo {volume} {33}},\ \bibinfo {pages} {117}
  (\bibinfo {year} {2017})}\BibitemShut {NoStop}%
\bibitem [{\citenamefont {Rem}\ \emph {et~al.}(2019)\citenamefont {Rem},
  \citenamefont {K{\"a}ming}, \citenamefont {Tarnowski}, \citenamefont
  {Asteria}, \citenamefont {Fl{\"a}schner}, \citenamefont {Becker},
  \citenamefont {Sengstock},\ and\ \citenamefont
  {Weitenberg}}]{rem2018identifying}%
  \BibitemOpen
  \bibfield  {author} {\bibinfo {author} {\bibfnamefont {B.~S.}\ \bibnamefont
  {Rem}}, \bibinfo {author} {\bibfnamefont {N.}~\bibnamefont {K{\"a}ming}},
  \bibinfo {author} {\bibfnamefont {M.}~\bibnamefont {Tarnowski}}, \bibinfo
  {author} {\bibfnamefont {L.}~\bibnamefont {Asteria}}, \bibinfo {author}
  {\bibfnamefont {N.}~\bibnamefont {Fl{\"a}schner}}, \bibinfo {author}
  {\bibfnamefont {C.}~\bibnamefont {Becker}}, \bibinfo {author} {\bibfnamefont
  {K.}~\bibnamefont {Sengstock}}, \ and\ \bibinfo {author} {\bibfnamefont
  {C.}~\bibnamefont {Weitenberg}},\ }\href@noop {} {\bibfield  {journal}
  {\bibinfo  {journal} {Nature Physics}\ }\textbf {\bibinfo {volume} {15}},\
  \bibinfo {pages} {917} (\bibinfo {year} {2019})}\BibitemShut {NoStop}%
\bibitem [{\citenamefont {Cao}\ \emph {et~al.}(2019)\citenamefont {Cao},
  \citenamefont {Tang}, \citenamefont {Guo}, \citenamefont {Chen},
  \citenamefont {Zhang},\ and\ \citenamefont {Zhou}}]{cao2019extraction}%
  \BibitemOpen
  \bibfield  {author} {\bibinfo {author} {\bibfnamefont {S.}~\bibnamefont
  {Cao}}, \bibinfo {author} {\bibfnamefont {P.}~\bibnamefont {Tang}}, \bibinfo
  {author} {\bibfnamefont {X.}~\bibnamefont {Guo}}, \bibinfo {author}
  {\bibfnamefont {X.}~\bibnamefont {Chen}}, \bibinfo {author} {\bibfnamefont
  {W.}~\bibnamefont {Zhang}}, \ and\ \bibinfo {author} {\bibfnamefont
  {X.}~\bibnamefont {Zhou}},\ }\href@noop {} {\bibfield  {journal} {\bibinfo
  {journal} {Optics Express}\ }\textbf {\bibinfo {volume} {27}},\ \bibinfo
  {pages} {12710} (\bibinfo {year} {2019})}\BibitemShut {NoStop}%
\bibitem [{\citenamefont {Erhard}(2004)}]{erhard2004experimente}%
  \BibitemOpen
  \bibfield  {author} {\bibinfo {author} {\bibfnamefont {M.}~\bibnamefont
  {Erhard}},\ }\emph {\bibinfo {title} {Experimente mit mehrkomponentigen
  Bose-Einstein-Kondensaten}},\ \href@noop {} {\bibinfo {type} {Phd thesis}},\
  \bibinfo  {school} {Universit\"{a}t Hamburg} (\bibinfo {year}
  {2004})\BibitemShut {NoStop}%
\bibitem [{\citenamefont {Kronj\"{a}ger}(2007)}]{kronjager2007coherent}%
  \BibitemOpen
  \bibfield  {author} {\bibinfo {author} {\bibfnamefont {J.}~\bibnamefont
  {Kronj\"{a}ger}},\ }\emph {\bibinfo {title} {Coherent Dynamics of Spinor
  Bose-Einstein Condensates}},\ \href@noop {} {\bibinfo {type} {Phd thesis}},\
  \bibinfo  {school} {Universit\"{a}t Hamburg} (\bibinfo {year}
  {2007})\BibitemShut {NoStop}%
\bibitem [{\citenamefont {Reinaudi}\ \emph {et~al.}(2007)\citenamefont
  {Reinaudi}, \citenamefont {Lahaye}, \citenamefont {Wang},\ and\ \citenamefont
  {Gu{\'e}ry-Odelin}}]{Reinaudi07}%
  \BibitemOpen
  \bibfield  {author} {\bibinfo {author} {\bibfnamefont {G.}~\bibnamefont
  {Reinaudi}}, \bibinfo {author} {\bibfnamefont {T.}~\bibnamefont {Lahaye}},
  \bibinfo {author} {\bibfnamefont {Z.}~\bibnamefont {Wang}}, \ and\ \bibinfo
  {author} {\bibfnamefont {D.}~\bibnamefont {Gu{\'e}ry-Odelin}},\ }\href@noop
  {} {\bibfield  {journal} {\bibinfo  {journal} {Optics letters}\ }\textbf
  {\bibinfo {volume} {32}},\ \bibinfo {pages} {3143} (\bibinfo {year}
  {2007})}\BibitemShut {NoStop}%
\bibitem [{\citenamefont {Hueck}\ \emph {et~al.}(2017)\citenamefont {Hueck},
  \citenamefont {Luick}, \citenamefont {Sobirey}, \citenamefont {Siegl},
  \citenamefont {Lompe}, \citenamefont {Moritz}, \citenamefont {Clark},\ and\
  \citenamefont {Chin}}]{hueck2017calibrating}%
  \BibitemOpen
  \bibfield  {author} {\bibinfo {author} {\bibfnamefont {K.}~\bibnamefont
  {Hueck}}, \bibinfo {author} {\bibfnamefont {N.}~\bibnamefont {Luick}},
  \bibinfo {author} {\bibfnamefont {L.}~\bibnamefont {Sobirey}}, \bibinfo
  {author} {\bibfnamefont {J.}~\bibnamefont {Siegl}}, \bibinfo {author}
  {\bibfnamefont {T.}~\bibnamefont {Lompe}}, \bibinfo {author} {\bibfnamefont
  {H.}~\bibnamefont {Moritz}}, \bibinfo {author} {\bibfnamefont {L.~W.}\
  \bibnamefont {Clark}}, \ and\ \bibinfo {author} {\bibfnamefont
  {C.}~\bibnamefont {Chin}},\ }\href@noop {} {\bibfield  {journal} {\bibinfo
  {journal} {Optics express}\ }\textbf {\bibinfo {volume} {25}},\ \bibinfo
  {pages} {8670} (\bibinfo {year} {2017})}\BibitemShut {NoStop}%
\bibitem [{\citenamefont {Song}\ \emph {et~al.}(2016)\citenamefont {Song},
  \citenamefont {He}, \citenamefont {Zhang}, \citenamefont {Hajiyev},
  \citenamefont {Huang}, \citenamefont {Liu},\ and\ \citenamefont
  {Jo}}]{Song:2016ep}%
  \BibitemOpen
  \bibfield  {author} {\bibinfo {author} {\bibfnamefont {B.}~\bibnamefont
  {Song}}, \bibinfo {author} {\bibfnamefont {C.}~\bibnamefont {He}}, \bibinfo
  {author} {\bibfnamefont {S.}~\bibnamefont {Zhang}}, \bibinfo {author}
  {\bibfnamefont {E.}~\bibnamefont {Hajiyev}}, \bibinfo {author} {\bibfnamefont
  {W.}~\bibnamefont {Huang}}, \bibinfo {author} {\bibfnamefont {X.-J.}\
  \bibnamefont {Liu}}, \ and\ \bibinfo {author} {\bibfnamefont {G.-B.}\
  \bibnamefont {Jo}},\ }\href@noop {} {\bibfield  {journal} {\bibinfo
  {journal} {Physical Review A}\ }\textbf {\bibinfo {volume} {94}},\ \bibinfo
  {pages} {061604} (\bibinfo {year} {2016})}\BibitemShut {NoStop}%
\end{thebibliography}%

\pagebreak
\newpage
\clearpage
\setcounter{equation}{0} \setcounter{figure}{0}
\renewcommand{\theequation}{S\arabic{equation}}
\renewcommand{\thefigure}{S\arabic{figure}}
\renewcommand{\thetable}{S\arabic{table}}

\onecolumngrid
\section*{\large\bf{Supplementary Materials}}

\begin{figure*}[!hb]
	\centering
	\includegraphics[width=0.6\linewidth]{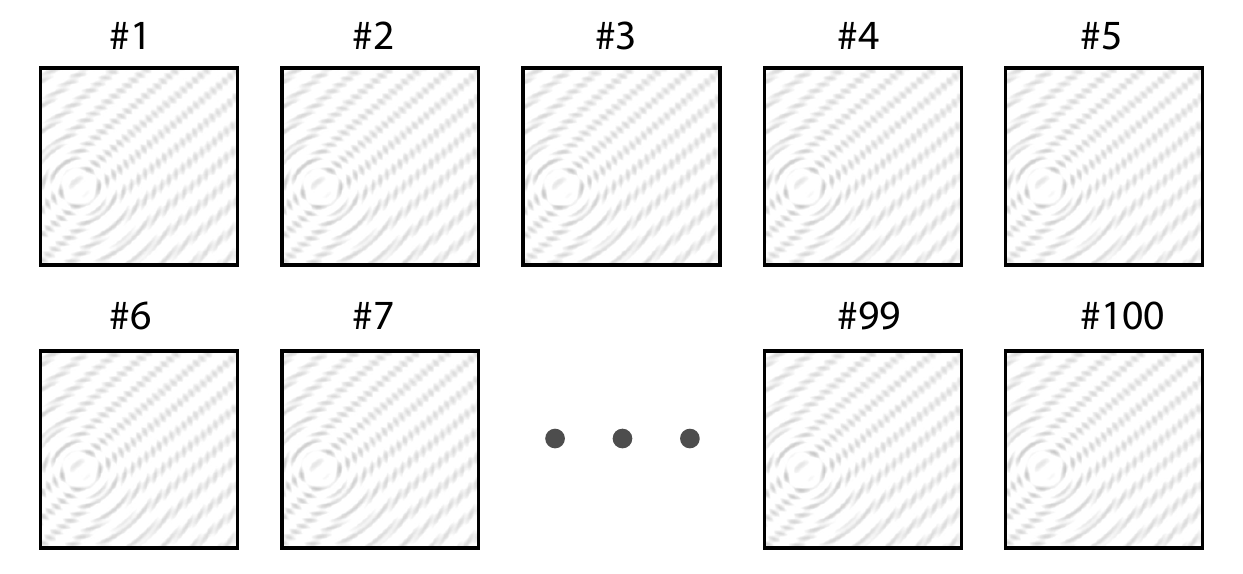}
	\caption{\textbf{Simulated fringe data} Images contains two types of fringes, concentric ring pattern and linear pattern respectively. The position of fringes are randomly shifted horizontally and vertically within 2 pixels.}
	\label{FigS1_SimulatedFringe}
\end{figure*}
\begin{figure*}[h]
	\centering
	\includegraphics[width=0.85\linewidth]{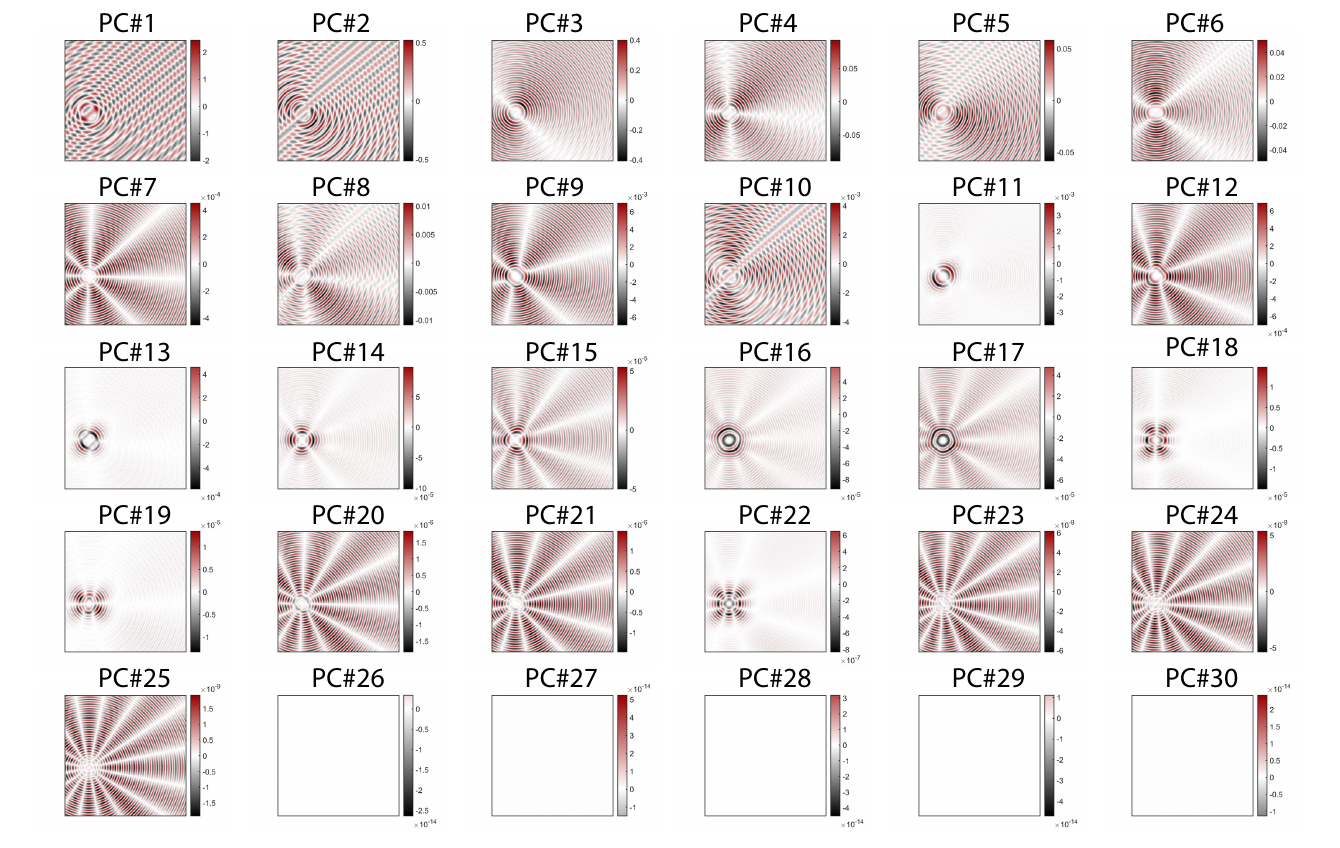}
	\caption{\textbf {Principal images (eigenvectors) of the simulated data set} First 30 principal images are shown.}
	\label{FigS2_SimulatedFringePCA}
\end{figure*}
\begin{figure*}
	\centering
	\includegraphics[width=0.85\linewidth]{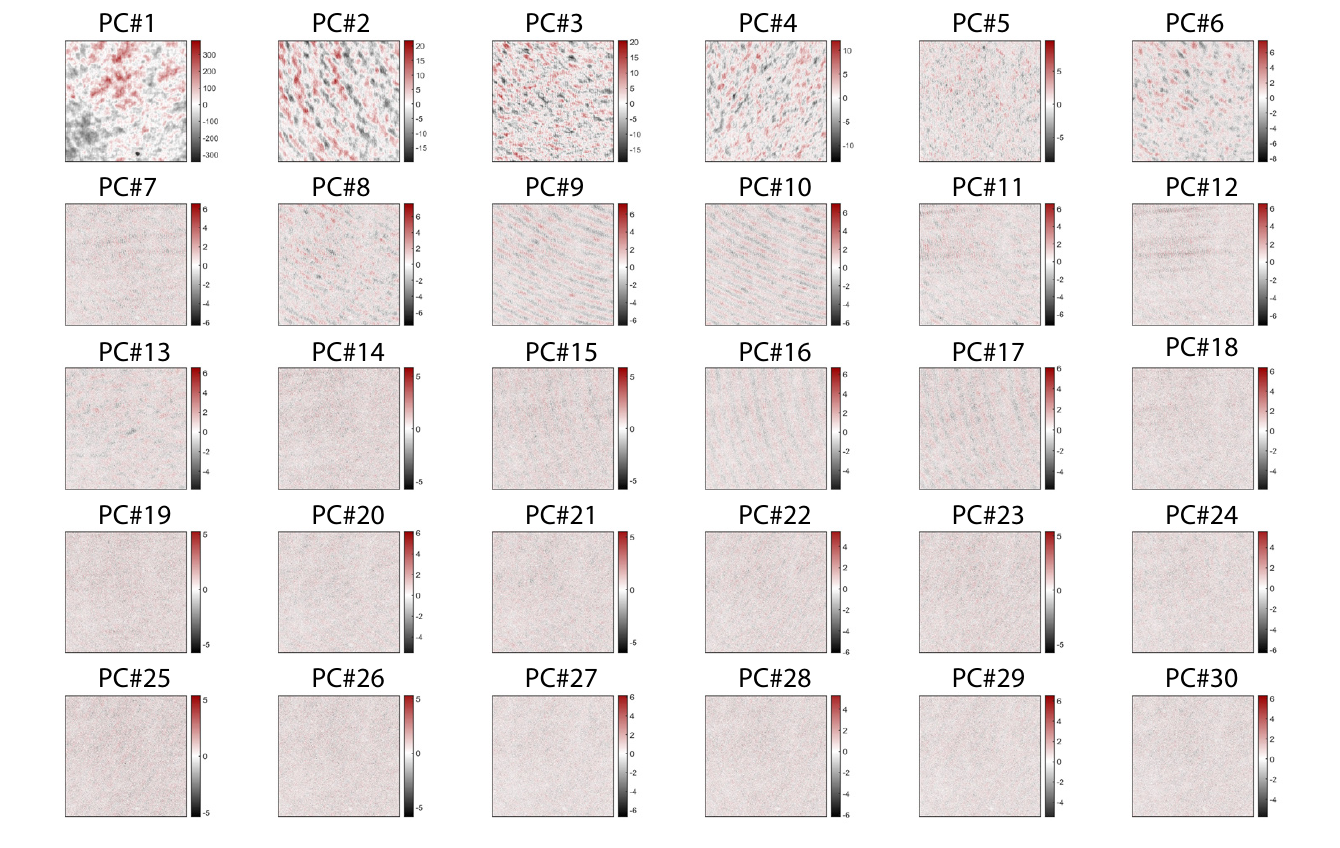}
	\caption{\textbf{Principal images (eigenvectors) of the measured reference image set} First 30 principal images are shown.}
	\label{FigS3_MeasFringePCA}
\end{figure*}

 We provide extended data sets that compliment the rank and the shift $d$ of both simulated and real image sets in the main text. We generate 100 images with two types of fringe patterns, linear and ring fringes shown in Fig.~\ref{FigS1_SimulatedFringe}. The fringes are randomly shifted within $d=2$ pixels both in the horizontal and vertical directions. Then PCA is applied to obtain the uncorrelated basis set. Fig.~\ref{FigS2_SimulatedFringePCA} shows the first 30 principal (component) images from the PCA result of this simulated data set. The significances of each principal images are plotted in Fig.\ref{Fig2_SetShift}(a) in the main text. Both the principal images and the significances reflect that the rank of the data set is around 25. The rank theoretically should be $(2\times d+1)^2=25$, as the shift has two degrees of freedom. Similarly, the Fig.~\ref{FigS3_MeasFringePCA} shows the principal images of real reference images taken by our imaging system, of which the significances are plotted in Fig.\ref{Fig2_SetShift}(b) in the main text. The rank is around 9 and thus we empirically choose the shift $d=1$ for the basis expansion due to $(2\times d+1)^2=9$.

\begin{enumerate}
	\item Figure \ref{FigS1_SimulatedFringe}. Simulated fringe data including the linear and ring fringes, are generated for Fig.\ref{Fig2_SetShift}(a) in the Main text.
	\item Figure \ref{FigS2_SimulatedFringePCA}. Principal images (eigenvectors) from the result of PCA applied to the simulated data set, of which significances are plotted in Fig.\ref{Fig2_SetShift}(a) in the Main text.
	\item Figure \ref{FigS3_MeasFringePCA}. Principal images (eigenvectors) from the result of PCA applied to the measured reference image set $\{I_i^{wo}\}$, of which significances are plotted in Fig.\ref{Fig2_SetShift}(b) in the main text.
\end{enumerate}


\end{document}